\documentclass[conference]{IEEEtran}
\IEEEoverridecommandlockouts
\usepackage{cite}
\usepackage{amsmath,amssymb,amsfonts}
\usepackage{mathtools}
\usepackage{algorithm}
\usepackage{algorithmic}
\usepackage{graphicx}
\usepackage{textcomp}
\usepackage{xcolor}
\usepackage{booktabs}
\def\BibTeX{{\rm B\kern-.05em{\sc i\kern-.025em b}\kern-.08em
    T\kern-.1667em\lower.7ex\hbox{E}\kern-.125emX}}
\begin{document}

\title{THFlow: A Temporally Hierarchical Flow Matching Framework for 3D Peptide Design}
% \vspace{-10mm}

\author{\IEEEauthorblockN{1\textsuperscript{st} Dengdeng Huang}
\IEEEauthorblockA{\textit{School of Computer Science} \\
\textit{Shanghai Jiao Tong University}\\
Shanghai, China \\
erika\_dd@sjtu.edu.cn}
\and
\IEEEauthorblockN{2\textsuperscript{nd} Shikui Tu}
\IEEEauthorblockA{\textit{School of Computer Science} \\
\textit{Shanghai Jiao Tong University}\\
Shanghai, China \\
tushikui@sjtu.edu.cn}
}

\maketitle

\begin{abstract}
Deep generative models provide a promising approach to de novo 3D peptide design. 
Most of them jointly model the distributions of peptide's position, orientation, and conformation, attempting to simultaneously converge to the target pocket. 
% However, there might be no physically meaningful conformational refinement for a randomly initialized peptide during the initial docking stage, as its atoms are too distant from the protein pocket to form a guiding interaction field. 
However, in the early stage of docking, optimizing conformation-only modalities such as rotation and torsion can be physically meaningless, as the peptide is initialized far from the protein pocket and no interaction field is present.
We define this problem as the multimodal temporal inconsistency problem and claim it is a key factor contributing to low binding affinity in generated peptides.
% primary reason for the low binding affinity and poor stability of the generated peptides. 
% To address this challenge, we propose THFlow, which explicitly models the temporal hierarchy between peptide position and conformation by employing a polynomial-based conditional flow to first accelerate the peptide’s position convergence. 
To address this challenge, we propose THFlow, a novel flow matching-based multimodal generative model that explicitly models the temporal hierarchy between peptide position and conformation. 
It employs a polynomial-based conditional flow to accelerate positional convergence early on, and later aligns it with rotation and torsion for coordinated conformation refinement under the emerging interaction field.
% It achieves this hierarchy by employing a polynomial-based conditional flow to first accelerate the peptide’s positional convergence.
Additionally, we incorporate interaction-related features, such as polarity, to further enhance the model's understanding of peptide-protein binding. 
Extensive experiments demonstrate that THFlow outperforms existing methods in generating peptides with superior stability, affinity, and diversity, offering an effective and accurate solution for advancing peptide-based therapeutic development.
\end{abstract}


\section{Introduction}
Peptides are short, single-chain proteins composed of amino acid.
Therapeutic peptide plays a pivotal role in applications, and their design is essential for achieving precise and stable docking with target receptor proteins~\cite{fosgerau2015peptide}.
The rapid development of deep learning has made 3D peptide design a feasible computational task, offering a powerful approach for developing therapeutics against protein-related diseases~\cite{muttenthaler2021trends, petsalaki2008peptide}.
Recent generative models tackle the vast and complex chemical space by integrating protein sequence and structure data to extract rich feature representations, facilitating the exploration of sequence-structure relationships~\cite{lin2023evolutionary}.
More recently, dynamic models such as diffusion and flow matching have gained prominence, simulating the physical docking process by generating trajectories for a peptide's movement via time-dependent vector fields~\cite{CIEMNY20181530, watson2023novo}. 
These models explicitly operate on key physical modalities, including displacement, rotation, and other conformational degrees of freedom. 
By directly modeling the evolution of these dynamics, they enable efficient exploration of conformational space and provide a realistic simulation of the docking process. 
Thus, they overcome the limitations of traditional experimental methods, such as high costs and long timelines~\cite{bhardwaj2016accurate}.

However, existing diffusion- and flow-based models struggle to generate peptides with both high binding affinity and structural stability~\cite{bennett2023improving}. 
One key limitation is their inability to model the temporal hierarchy across different modalities. 
Although these models generate multimodal variables over discrete timesteps, they often assume that all modalities evolve in a strictly synchronized fashion~\cite{corso2022diffdock, li2024full, lin2024ppflow}. 
In reality, peptides must first move toward the target binding pocket, and only then undergo fine-grained conformational adjustments, such as backbone rotation and side-chain torsion, in response to the local interaction environment~\cite{london2013peptide}. 
As a result, at each simulation timestep, the peptide’s position, conformation, and amino acid type are combined in a temporally misaligned manner, violating the natural progression of peptide dynamics. 
This multimodal temporal inconsistency problem leads to inaccurate modeling of peptide dynamics and reduces the biological plausibility of the simulation.

\begin{figure}[ht]
\vspace{-1mm}
    \begin{center}
    \includegraphics[width=0.98\linewidth]{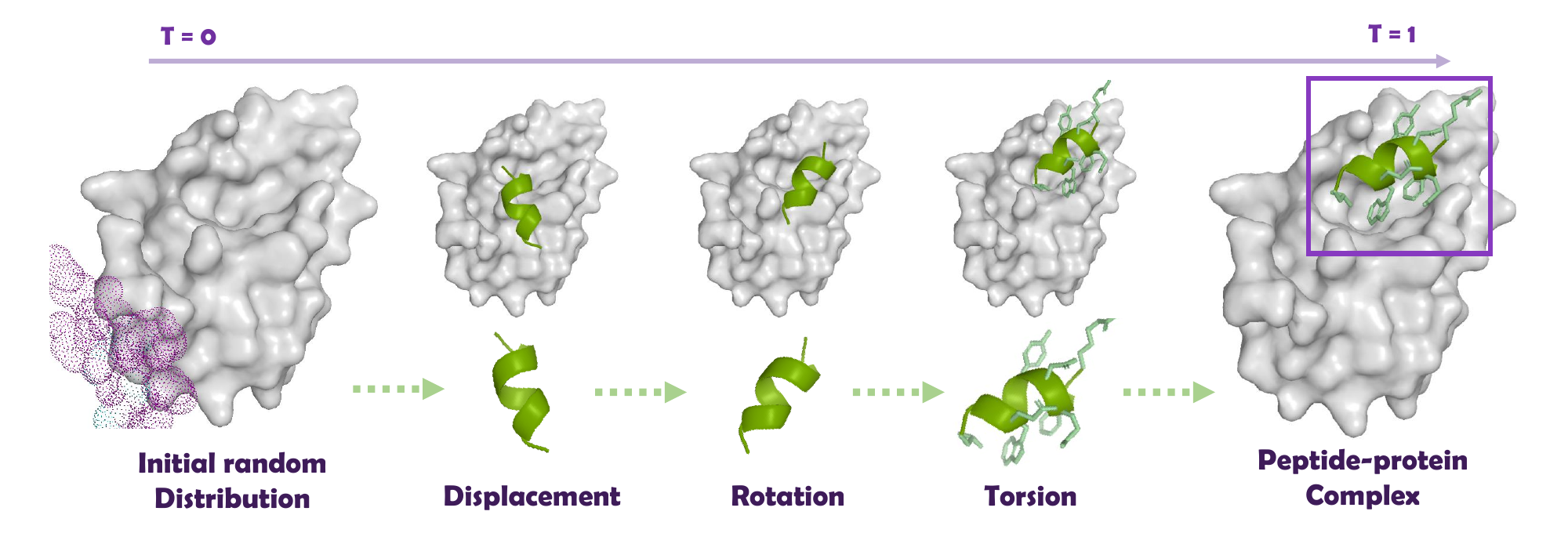}
    \vspace{-2mm}
    \caption{Temporally hierarchical peptide docking process}
    % : Position transformations occur prior to rotational and torsional changes due to the requirement that the peptide first moves into proximity with the target pocket, allowing for residue interactions with the surrounding environment.}
    \label{icml-historical}
    \end{center}
\vspace{-1mm}
\end{figure}

To address this multimodal temporal inconsistency problem, we propose \textbf{THFlow}, a novel \textbf{T}emporally \textbf{H}ierarchical multimodal \textbf{flow} matching method.
It accelerates the peptide’s movement toward the target binding pocket, followed by fine-tuning its conformation and posture, closely mirroring the actual docking process. 
To realize this, the core idea is to design a time-varying flow for the position modality, where the flow’s gradient vector decreases in magnitude over time, accelerating convergence in the initial stage.
Meanwhile, the other modalities converge at a constant rate, effectively decoupling the evolution of position from conformation and posture, thus capturing the temporally hierarchical nature of the different modalities.
Additionally, to prevent the model from focusing solely on conformational alignment while neglecting docking interactions, interaction-related information such as polarity, charge, and hydrophilicity is incorporated, ensuring stronger binding affinity to the target protein.

The THFlow framework consists of two key components:
(i) During the pre-training phase, the model learns the evolution of four modalities: peptide position, rotation, torsion, and amino acid types, from noise to the target distribution. 
A neural network is used to independently learn the gradient dynamics for each modality.
(ii) In the sampling phase, based on ordinary differential equations (ODEs)~\cite{teschl2021ordinary}, the pre-trained model predicts the gradient vector for the next timestep based on the current timestep's information, progressively simulating the peptide docking process.

THFlow provides two key advantages: 
(i) it introduces a time-varying flow to decouple the convergence of different modalities, transforming the evolution from synchronized to asynchronous across temporal hierarchies. 
This enables more accurate peptide docking simulations, resulting in the generation of \textbf{twice as many stable peptide-protein complexes} compared to the baseline; 
(ii) the rapid convergence of the position modality reduces ineffective time spent on conformation exploration in the early stages, allowing the model to fine-tune conformation and posture near the binding pocket. 
By incorporating interaction-related information, the model explores interaction forces in greater depth, leading to a \textbf{7.13\% increase in the affinity} of the generated peptides.

In summary, our key contributions include: 
\begin{itemize}
    \item We first identify the \textbf{multimodal temporal inconsistency problem} as a critical issue in peptide design models, which violates biological principles and hinders the generation of high-affinity peptides.
    \item We design a novel temporally hierarchical flow matching framework \textbf{THFlow} that decouples the evolution of different modalities, providing a more accurate simulation of docking process.
    \item Extensive experiments demonstrate THFlow’s superior performance in both co-design and re-docking, significantly improving generated peptides' \textbf{stability} and \textbf{affinity}, while providing an efficient method for advancing peptide-based therapeutic development.
\end{itemize}

\section{Related Work}
\subsection{Peptide generation}
Peptide generation is a complex task in drug discovery, aiming to design peptides with specific biological functions, such as binding to target proteins or modulating enzymatic activity. 
Generative models have significantly advanced peptide design by exploring large sequence and structural spaces with greater flexibility. 
For example, AMP-diffusion~\cite{chen2024amp} utilizes the capabilities of protein large language model ESM-2~\cite{beal2015esm} to regenerate functional antimicrobial peptides (AMPs). 
MMCD~\cite{wang2024multi} employs multi-modal contrastive learning in diffusion, exploiting the integration of both sequence and structural information to produce peptides with high functional relevance.
Besides, RFDiffusion~\cite{watson2023novo}, created for protein design, inspired the application of similar diffusion techniques for peptide generation. 

Peptide design is a subtask of peptide generation, involving several key components.
One important aspect is backbone design, where methods like PepFlow~\cite{li2024full} and PPFlow~\cite{lin2024ppflow} use flow matching to simulate dynamic conformational changes and optimize peptide properties. 
Another aspect is side-chain packing, with methods such as RED-PPI~\cite{luo2023rotamer} focusing on protein-protein complexes and DiffPack~\cite{zhang2024diffpack} targeting peptide-protein interactions.

\subsection{Protein–ligand docking}
Protein–ligand docking aims to predict the binding pose and affinity of a ligand to a target protein. 
Traditional docking methods often rely on rigid-body simulations and predefined scoring functions, which can struggle to handle flexible ligands and complex interactions~\cite{totrov1997flexible}.
Recent years, deep learning (DL) approaches have significantly improved docking accuracy. 
Models like AtomNet~\cite{stafford2022atomnet} and OnionNet~\cite{wang2021onionnet} use convolutional neural networks (CNNs) to capture complex molecular features, significantly enhancing binding affinity predictions.
Further developments have incorporated graph-based models, such as GraphSite~\cite{shi2022graphsite} and DGraphDTA~\cite{yang2022mgraphdta}, which represent protein-ligand interactions as graphs to better model flexible docking. 
Additionally, advances in protein structure characterization, notably via AlphaFold~\cite{evans2021protein}, have enhanced understanding of protein flexibility, fueling the development of structure-based docking prediction methods~\cite{johansson2022improving}.
However, challenges still remain in fully capturing the dynamic and flexible nature of ligands, highlighting the need for further improvements in docking models.

\section{Method}
\subsection{Preliminary}
\subsubsection{Problem Statement}
A peptide is a short sequence of amino acids, where each amino acid has a specific type, denoted by \( a \), and consists of a rigid backbone and a flexible side chain.
\begin{figure}[!h]
    \centering
    \vspace{-2mm}
    \includegraphics[width=0.7\linewidth]{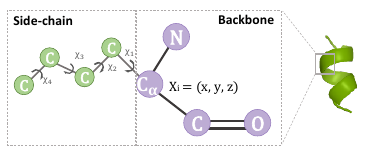}
    \vspace{-1mm}
    \caption{Rigid peptide backbone and flexible side chains.}
    \vspace{-2mm}
    \label{fig:pep}
\end{figure}

The backbone consists of four heavy atoms \( N \), \( C_{\alpha} \), \( C \), and \( O \), with the position of \( C_{\alpha} \) serving as the reference, denoted as \( \mathbf{X}_i \) at the origin.
Attached to this alpha-carbon \( C_{\alpha} \) is the side chain, a variable chemical group that determines its unique properties.
Peptide's orientation is modeled in two parts: the overall rigid rotation, represented by the rotation matrix \( \mathbf{R} \in \operatorname{SO}(3) \), and the side-chain torsion, captures four torsion angles \(\boldsymbol{\tau} = (\chi_1, \chi_2, \chi_3, \chi_4)\) of the rotatable bonds.

In the docking process, the peptide acts as the ligand, represented as \( Lig = \{ ( \mathbf{X}_i, \mathbf{R}_i, \boldsymbol{\tau}_i, \boldsymbol{a}_i) \}_{i=1}^n \), where $n$ is the length of peptide, while the protein serves as the receptor, denoted as \( Rec \). 
The task of designing the peptide \( Lig \) based on a spesific \( Rec \) can be formulated as learning the conditional distribution \( p(Lig | Rec) \). 
Expanding the problem to the four modalities we have modeled, the problem can be empirically decomposed into the product of probabilities of the four components: 
\begin{align}
    p(Lig | Rec) \propto &p(\{\mathbf{X}_i\}_{i=1}^n | Rec) \cdot p(\{\mathbf{R}_i\}_{i=1}^n | Rec) \nonumber\\
    \cdot &p(\{\boldsymbol{\tau}_i\}_{i=1}^n | Rec) \cdot p(\{\boldsymbol{a}_i\}_{i=1}^n | Rec).
\end{align}

\subsubsection{Conditional Flow Matching}
Conditional Flow Matching (CFM)~\cite{lipman2022flow} models the transformation between a prior distribution \( p = p_0 \) and a target data distribution \( q = p_1 \) on a manifold \( \mathcal{M} \) using ODEs. 
Mathematically, this ``flow'' represents a probability path generated by a time-dependent vector field.
The transformation is described as:
\begin{equation}
    \frac{d}{dt} \varphi_t(x) = u_t(x),
\end{equation}
where \( \varphi_t(x) \) represents the data distribution at time \( t \), and \( u_t(x) \) is the time-dependent vector field that driving the flow. 
% Since the true vector field \( u_t(x) \) is unknown, it is assumed based on a hypothesis form and a known target distribution \( x_1 \).
% Usually a neural network is used to approximate this flow by learning a vector field \( v_t(x) \). 
In the flow matching framework, the true vector field \( u_t(x) \) is considered intractable. 
Therefore, the method defines a conditional vector field \( u_t(x|x_1) \) based on a target data point \( x_1 \) to serve as a tractable learning target. 
A neural network is then trained to approximate this conditional field, and its output is denoted as the learned vector field \( v_t(x) \).
% The loss is defined to minimize the difference between \( v_t(x) \) and \( u_t(x|x_1) \):
The model is trained to minimize the difference between its predicted vector field \( v_t(x) \) and the conditional target field \( u_t(x|x_1) \) via the CFM loss:
\begin{equation}
    L_{\text{CFM}}(\theta) = \mathbb{E}_{t, p_t(x|x_1)} \| v_t(x) - u_t(x|x_1) \|^2,
\end{equation}

In most current methods, the distribution \( \varphi_t(x) \) is typically assumed to follow a simple linear interpolation between the initial distribution \( x_0 \) and target distribution \( x_1 \): \( \varphi_t(x) = t x_1 + (1 - t) x_0 \), where \( t \in [0, 1] \).
The corresponding vector field is \( u_t(x|x_1) = \frac{d}{dt} \varphi_t(x) = x_1 - x_0 \).
This form has been proven to correspond to the Optimal Transport (OT) solution~\cite{lipman2022flow}, which provides an efficient and fast means of transforming distributions with a fixed direction and magnitude.

\subsubsection{Time-varying Flow and Diffusion Formulation}
Both diffusion and flow matching model the transformation of distributions as an evolving path \( \varphi_t(x) \), which can be defined as an interpolation of the target signal and the initial noise:
\begin{equation}
\varphi_t(x) = \mu(t)x_1 + \sigma(t)x_0.
\end{equation}

Here, \( \mu(t) \) and \( \sigma(t) \) are time-dependent scalar coefficients satisfying the boundary conditions \( \mu(0)=0, \sigma(0)=1 \) and \( \mu(1)=1, \sigma(1)=0 \). 

In diffusion models, the transformation is typically defined by variance-preserving stochastic differential equations (SDEs)~\cite{croitoru2023diffusion}. 
This corresponds to a constraint where the sum of squares of the coefficients is one: \( \mu(t)^2 + \sigma(t)^2 = 1 \). 
This formulation maintains signal stability by ensuring that the variance of the evolving distribution remains constant throughout the transformation.

In contrast, flow matching under the OT assumption defines a direct, linear path. 
It is realized with \( \mu(t) = t \) and \( \sigma(t) = 1-t \), which corresponds to a simpler linear constraint where the coefficients sum to one: \( \mu(t) + \sigma(t) = 1 \). 
This linear structure provides an intuitive and efficient straight-line trajectory from noise to data.

Building on this, our proposed time-varying flow corresponds to a polynomial interpolation: \( \mu(t) = 1 - (1 - t)^k \) and \( \sigma(t) = (1 - t)^k \), which also adheres to the linear sum constraint \( \mu(t) + \sigma(t) = 1 \), and reduces to the linear flow when \( k = 1 \). 
For \( k > 1 \), the flow provides increased flexibility to simulate the varying convergence speed of the position manifold in docking tasks, accurately capturing the complex dynamics of peptide conformation changes.

\subsection{Time-varying Flow for Position}
\label{time-varying flow}
Position manifold refers to the coordinates of the C-alpha atoms of the peptide backbone in Euclidean space. 
We initialize the system using a standard Gaussian distribution \( \mathbf{X}^0 \sim \mathcal{N}(0, I) \) as the random initialization, with \( \mathbf{X}^1 \) representing the target distribution. 
The task is to model the trajectory from \(  \mathbf{X}^0 \) to \( \mathbf{X}^1 \). 

Despite being efficient and fast, the OT assumption applied across all modalities leads to synchronized evolution in time.
To resolve this multimodal temporal inconsistency problem, we define a polynomial interpolation for position mainfold to explicitly decouples its evolution from other mainfolds. 
The interpolation is defined as:
\begin{equation}
    \mathbf{X}_i^t = (1 - t)^k\mathbf{X}_i^0 + (1 - (1 - t)^k) \cdot \mathbf{X}_i^1, \quad t \in [0, 1],
\end{equation}
where $k$ is a hyperparameters.

Taking the derivative of this equation with respect to time $t$ gives the corresponding gradient vector field: 
\begin{equation}
    u_t^{\mathrm{pos}}(X_i^t|X_i^0, X_i^1) = \frac{d\mathbf{X}_i^t}{dt} = k(1 - t)^{k-1} (\mathbf{X}_i^1 - \mathbf{X}_i^0),
\end{equation}
From this formulation, we can see that the gradient vector field $u_t^{\mathrm{pos}}$ is time-varying and decreases over time, while the direction remains consistent, always aligned with \( \mathbf{X}_i^1 - \mathbf{X}_i^0 \). 
In a broader view, the position manifold evolves quickly towards \( \mathbf{X}^1 \) near the initial state, reflecting the initial stage of docking where the peptide rapidly moves towards the pocket. 
As the docking progresses, the evolution slows down, mirroring the peptide's fine-tuning of its conformation near the pocket, including rotation and torsion adjustments.
The polynomial flow resolves the temporal consistency limitation of synchronized OT assumptions, enabling both efficient exploration of conformational space and precise optimization of flexible docking.
% \vspace{-5mm}

\paragraph{Loss Function for Position Flow}
We use a trainable neural network \( v^\mathrm{pos} \) to approximate the time-varying gradient \( u_t^\mathrm{pos} \). 
The loss function is designed to minimize the squared error between the predicted and target gradient vectors over all timesteps, ensuring that the model learns the correct temporal dynamics. 
It can be expressed as:
\begin{equation}
\small
    \mathcal{L}_i^{\mathrm{pos}} = \mathbb{E}_{\substack{p(X_i^0), p(X_i^1)\\ p(X_i^t | X_i^0, X_i^1)}} \left\| v_t^\mathrm{pos}(X_i^t) - k(1 - t)^{k-1}(X_i^1 - X_i^0) \right\|^2.
\end{equation}

\begin{figure*}
    \centering
    \vspace{-2mm}
    \includegraphics[width=0.99\linewidth]{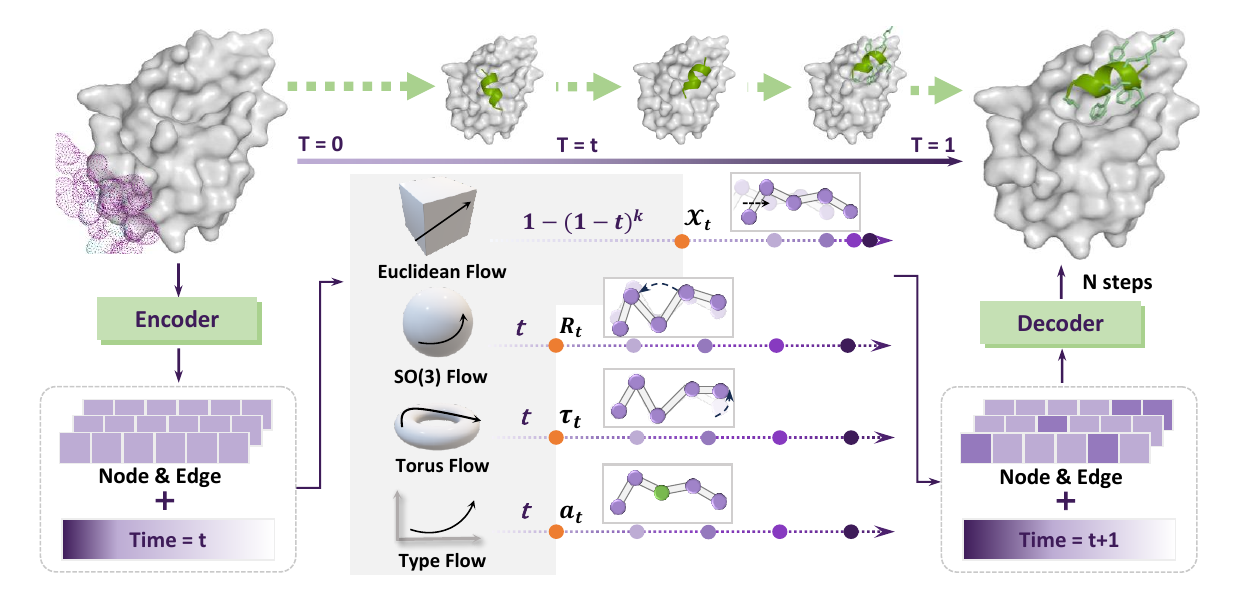}
    \vspace{-1mm}
    \caption{Overview of the THFlow framework}
    \label{fig:overview}
    \vspace{-2mm}
\end{figure*}

\paragraph{Non-Uniform Time Sampling Strategy}  

To address the bias of standard uniform sampling—where interpolated positions \( \mathbf{X}^t \) cluster near \( \mathbf{X}^1 \) under polynomial-based interpolation—we redefine the time variable as: \(t = z^k, z \sim \mathcal{U}(0,1)\).
This non-uniform sampling adjusts the temporal density, prioritizing the initial phase (\( t \to 0 \)) where rapid positional convergence occurs. 
By reshaping the probability distribution of \( t \), the strategy ensures balanced coverage of the entire trajectory, preventing under-sampling of critical early dynamics. 
Simultaneously, it allocates more training samples to the information-sparse initial phase, enabling the model to robustly learn fast positional alignment governed by large initial gradients, while stabilizing gradient updates across all stages. 
% This harmonizes sampling with the peptide's hierarchical docking process: rapid approach first, refinement later.

\subsection{Linear Flow for Orientation and Amino Acid Type}
In this section, we follow recent advances in flow-based methods~\cite{lin2024ppflow, li2024full}, modeling the flows of rotation, torsion, and amino acid types using the optimal transport assumption, which enables efficient transitions between states.
% As the peptide approaches the pocket, these manifolds adjust according to the pocket environment, which ensures temporal consistency and allows us to assume synchronous convergence.
As the peptide approaches the pocket, these manifolds adjust according to the pocket environment, ensuring temporal consistency. 
This adjustment allows us to treat these three modalities as time-synchronized linear flows. 
Together with the time-varying flow model discussed in the previous section, our approach decouples the evolution of the position manifold from the other manifolds, effectively capturing cross-modal temporal inconsistencies.

\subsubsection{Rotation Matrices}
A rotation matrix, \( \mathbf{R}_i \), is an element of the special orthogonal group \( SO(3) \), which describes rigid body rotations. As a Lie group, \( SO(3) \) enables its elements to be locally represented by its tangent space. It has been proven that the exponential map facilitates smooth interpolation in Lie groups. Therefore, we use the rotation matrices under the exponential map for linear interpolation, which can be expressed as:
\begin{equation}
    R_i^t = \exp_{R_0^i} \left( t \log_{R_0^i}  \left( R_1^i \right) \right),
\end{equation}
\begin{equation}
    u_t^\mathrm{rot}(R_i^t|R_i^0,R_i^1)=\frac{\log_{R_i^t}R_i^1}{1-t},
\end{equation}
The loss function \( \mathcal{L}_i^{\mathrm{rot}} \) is defined as:
\begin{equation}
    \mathcal{L}_i^{\mathrm{rot}} = \mathbb{E}_{\substack{p(R_i^0), p(R_i^1) \\ p(R_i^t | R_i^0, R_i^1)}} \left\| v_t^\mathrm{rot}(R_i^t) - \frac{\log_{R_i^t} R_i^1}{1-t} \right\|_{\mathrm{SO}(3)}^2.
\end{equation}

\subsubsection{Torsion Angles}  
The torsion angle, \( \boldsymbol{\tau}_i \in [0, 2\pi)\), refers to the angle of rotation between two planes formed by four consecutive atoms in the side-chain of a peptide. 
Unlike the rotation of a rigid body, torsion alters the internal structure without changing the overall properties. 
The torsion angle has a periodicity of \( 2\pi \), and its space is topologically a torus, with the angle smoothly wrapping around itself as it changes continuously. 
Therefore, we define \( \boldsymbol{\tau}_i^t \):
\begin{equation}
    \boldsymbol{\tau}_i^t = (1-t) \boldsymbol{\tau}_i^0 + t \boldsymbol{\tau}_i^1 \mod 2\pi,
\end{equation}
\begin{equation}
    u_t^\mathrm{toru}(\boldsymbol{\tau}_i^t|\boldsymbol{\tau}_i^0,\boldsymbol{\tau}_i^1) = \boldsymbol{\tau}_i^1 - \boldsymbol{\tau}_i^0 \mod 2\pi,
\end{equation}
The loss function \( \mathcal{L}_i^{\mathrm{toru}} \) is defined as:
\begin{equation}
    \mathcal{L}_i^{\mathrm{toru}} = \mathbb{E}_{\substack{p(\boldsymbol{\tau}_i^0), p(\boldsymbol{\tau}_i^1) \\ p(\boldsymbol{\tau}_i^t | \boldsymbol{\tau}_i^0, \boldsymbol{\tau}_i^1)}} \left\| v_t^\mathrm{toru}(\boldsymbol{\tau}_i^t) - (\boldsymbol{\tau}_i^1 - \boldsymbol{\tau}_i^0 \right)\|^2.
\end{equation}

\subsubsection{Amino Acid Types}  
% The residue type of peptide, \( a_i \in {1,2...,20}\), is a discrete variable representing the identity of an amino acid in the peptide sequence. To smooth this representation, we treat the residue types as continuous logits in a 20-dimensional space. After applying soft one-hot encoding to the residue types, the logits are linearly interpolated between the initial type \( s_i^0 \) and the target type \( s_i^1 \). 
A peptide is composed of a sequence of amino acids, where the amino acid type at position \( i \), \( a_i \in \{1, 2, ..., 20\} \), has 20 possible distinct values. Since \( a_i \) is a discrete variable, we define the soft label \( s_i \) as continuous logits in a 20-dimensional space to facilitate smooth linear interpolation of this manifold.
The interpolation for soft label \( s_i \) is defined as:
\begin{equation}
    s_i^t = t s_i^1 + (1 - t) s_i^0,
\end{equation}
\begin{equation}
    u_t^\mathrm{type}(s_i^t|s_i^0, s_i^1) = s_i^1 - s_i^0,
\end{equation}
The loss function \( \mathcal{L}_i^{\mathrm{type}} \) is defined as:
\begin{equation}
    \mathcal{L}_i^{\mathrm{type}} = \mathbb{E}_{\substack{p(s_i^0) , p(s_i^1)\\ p(s_i^t | s_i^0, s_i^1)}} \left\| v_t^\mathrm{type}(s_i^t) - (s_i^1 - s_i^0) \right\|^2.
\end{equation}

\subsection{Loss Balancing for Sequence-Structure Co-design}
To balance sequence and structural alignment, we define the spatial loss $\mathcal{L}^{\mathrm{spa}}$ as the weighted sum of the position, rotation and torsion losses:
\begin{equation}
\mathcal{L}^\mathrm{spa} = \sum_{l \in \{\text{pos}, \text{rot}, \text{tor}\}} \alpha_l \cdot \mathcal{L}^l,
\end{equation}
where $\alpha_l$ are the corresponding hyperparameters.

The sequence loss weight, $\alpha_{\mathrm{type}}$, is dynamically adjusted based on the spatial loss:
\begin{equation}
\alpha_{\text{type}} = \min\left( \max\left( \frac{20}{\mathcal{L}^\mathrm{spa}}, 1 \right), \alpha_{\text{type}}^{max} \right),
\end{equation}
where $\alpha_{\mathrm{type}}^{\text{max}}$ is a hyperparameter that controls the upper bound. 
This ensures that when spatial loss is small, indicating structural alignment with the reference, the sequence is encouraged to match it; otherwise, the alignment requirement is relaxed.
The total loss is then computed as:
\begin{equation}
\mathcal{L}^{total} = \sum_{l \in \{\text{pos}, \text{rot}, \text{tor}, \text{type}\}} \alpha_l \cdot \mathcal{L}^l.
\end{equation}

\subsection{Sampling with ODE}
% \paragraph{}
We perform sampling with the pre-trained flow model by formulating peptide generation as an ODE, where the peptide ${Lig}$ evolves according to $\frac{d}{dt} Lig = v_t({Lig})$, with derivatives computed for each of the four modalities. 
The equation is discretized into $N$ steps, yielding the final peptide $\overline{}{{Lig}_1}$.

Each modality's update depends not only on its own state but also on the states of other modalities. 
This interdependence highlights the importance of accurately capturing the temporal relationships across modalities. 
By iteratively updating these states, our model effectively simulates the dynamic nature of peptide docking, while enabling flexible sampling strategies for different design tasks such as fix sequence for re-docking, and fix backbone for side-chain packing.

\begin{algorithm}[!htb]
    \caption{Sampling with ODE}
    \begin{algorithmic}
        \STATE {\bfseries Input:} $\text{Enc}(Rec)$
        \STATE {\bfseries Init:} Initial state $Lig^0 = \{(X_i^0, R_i^0, \boldsymbol{\tau}_i^0, a_i^0)\}_{i=1}^{n}$
        \FOR{$t = 1$ {\bfseries to} $N$}
            \STATE Predict gradient vecto $v_t = \text{Dec}(Lig^{\frac{t-1}{N}}, Rec, t)$
            \STATE EulerStep ${Lig^{\frac{t}{N}}} = Lig^{\frac{t-1}{N}} + \Delta t \cdot v$
        \ENDFOR
        \STATE {\bfseries Output:} $\overline{{Lig}^1}$
    \end{algorithmic}
\end{algorithm}

\subsection{Network Parametrization}
\subsubsection{Encode with Interaction-related Information}
We use two multi-layer perceptrons (MLPs) to encode the features of amino acids and the relationships between residue pairs.
The first MLP encodes the features of individual residues, processing the amino acid type, backbone dihedral angles, and local atomic coordinates. 
Position embeddings are integrated to identify the context of  sequences.
Additionally, properties such as polarity, charge, hydrophilicity, and sulfur presence, which is critical for stability, are encoded using one-hot vectors, serving as auxiliary interaction-related information. 
This allows the model to not only focus on structural alignment but also on the alignment of interaction forces, ultimately enhancing the affinity between the peptide and protein.
The second MLP encodes the relationships between residue pairs, capturing their relative positions, distances between atoms, and dihedral angles.
% \vspace{-3mm}
\subsubsection{Learning Gradient Vector with IPA and Transformers}  
We use invariant point attention (IPA)~\cite{lee2019set} and transformer to learn the gradient vector field \( v_t \). 
The model consists of multiple IPA blocks for spatial feature learning, followed by transformer encoders for sequence modeling. 
Each block includes layer normalization, transition layers, and backbone updates, iteratively refining peptide conformations. 
Residue identity, angular encoding, and time-step embedding are integrated to ensure smooth temporal updates.

\vspace{1mm}

\section{Experiment}
\subsection{Set up}
In this section, we evaluate THFlow on four tasks: (i) sequence-structure co-design, (ii) re-docking, (iii) side-chain packing and (iv) one-step generation.
Through these evaluations, we assess whether the correct temporal characterization leads to the generation of more stable, high-affinity peptides with accurate structural alignment and functional relevance.

% \subsection{Dataset}
The dataset, derived from the work of PepFlow~\cite{li2024full}, was obtained from PepBDB~\cite{wen2019pepbdb} and Q-BioLip~\cite{wei2024q}.
To ensure high-quality data, duplicates were removed, a resolution threshold of less than 4 \AA was applied, and peptide lengths were restricted to between 3 and 25 residues.
This preprocessing resulted in a final dataset consisting of 10,348 complexes, with 166 complexes reserved for the test set, and the remaining data split for training and validation.

% \subsection{Set up}
During pre-training, we simply selected the hyperparameter \( k = 2 \) for the time-varying flow model discussed in section~\ref{time-varying flow}. 
We trained three variants of our model to evaluate different configurations: 
(i) \textbf{THFlow}, the full model with all components.
(ii) \textbf{THFlow \textit{w/o} II}, which excludes the interaction-force related information.
(iii) \textbf{THFlow \textit{w/o} II+LW}, which further removes the sequence-structure balance weight.
% Each model variant was trained for a total of 70,000 iterations to ensure sufficient training and convergence. The experiments were conducted on a Tesla P40 GPU with a batch size of 12, providing consistent conditions for model evaluation.
Each model variant was trained for a total of 65,000 iterations to ensure sufficient training and convergence. The experiments were conducted on a Tesla P40 GPU with a batch size of 12, using the Adam optimizer with a learning rate of \(5 \times 10^{-4}\) and a plateau learning rate scheduler with a factor of 0.8, patience of 10, and a minimum learning rate of \(5 \times 10^{-6}\).

\subsection{Sequence-Structure Co-design}
Sequence-structure co-design involves jointly generating the peptide sequence and conformation, resulting in a full-atom peptide docked onto the target protein.
% \paragraph{Baseline Model}
We evaluate three baseline models: \begin{sc}DiffPP\end{sc}, \begin{sc}PPFlow\end{sc}~\cite{lin2024ppflow} and \begin{sc}PepFlow\end{sc}~\cite{li2024full}. 
\begin{sc}DiffPP\end{sc} is a diffusion model for protein backbone parametrization, using DDPM~\cite{yang2023diffusion} and SO(3)-DPM~\cite{leach2022denoising} to model translation and rotation, along with multinomial diffusion for amino acid types. 
\begin{sc}PepFlow\end{sc} and \begin{sc}PPFlow\end{sc}, which were proposed simultaneously, are the latest models based on the flow matching framework for peptide design. They represent peptide structures by modeling backbone frames on the SE(3) manifold and side-chain dynamics on high-dimensional tori, enabling the generation of full-atom peptides with a focus on structural accuracy and torsion angle optimization.

\begin{table}[!h]
\vspace{-1mm}
\caption{Comparison for Sequence-Structure Co-design: \textbf{Highest Affinity and Stability}. \textit{Stab.} $=$ Stability, \textit{Aff.} $=$ Affinity, \textit{Nov.} $=$ Novelty, \textit{Div.} $=$ Diversity}
\label{co-design}
% \vspace{-1mm}
    \begin{center}
    % \begin{small}
    % \begin{sc}
        \begin{tabular}{ccccccr}
        \toprule
        \textbf{Method}   & \textbf{Stab. $\uparrow$}& \textbf{Aff. $\uparrow$}&\textbf{Nov. $\uparrow$}& \textbf{Div.}&\textbf{AAS}\\
        \midrule
        PepFlow      & 4.16\%  & 13.19\% &50.45\% & 0.461 &  \textbf{62.94\%}\\
        PPFlow      & 3.05\%     & 10.68\%      & 85.07\% & \textbf{0.705}    & 25.40\% \\
        DiffPP      & 5.34\%     & 12.21\%      & 79.15\%  & 0.688     & 25.83\% \\
        \midrule
        THFlow  & \textbf{12.33\%} & \textbf{17.81\%} & 84.94\% & 0.651 & 44.61\% \\
        THF \textit{w/o} II & 10.42\% & 11.11\% &85.34\% & 0.648 &  48.26\% \\
        THF \textit{w/o} II+LW &4.32\% & 7.19\% &\textbf{86.23\%} & 0.645 &  49.27\% \\
        \toprule
        \end{tabular}
    % \end{sc}
    % \end{small}
    \end{center}
% \vskip -0.1in
\vspace{-2mm}
\end{table}

% \vspace{-2mm}
\subsubsection{Metrics}
The evaluation of the generated peptides is based on five metrics. Energy is calculated using Rosetta~\cite{rohl2004protein}, with two primary measures: \textbf{Affinity}, which quantifies the binding energy \(\Delta G\), where lower values indicate stronger binding potential, and \textbf{Stability}, representing the overall energy of the peptide-protein complex, with lower values indicating a more stable complex. 
We report the percentages of designed peptides with higher affinity and stability than the reference ones. 
\textbf{Diversity} is evaluated using the TM-score~\cite{zhang2005tm}, defined as the average of 1 minus the pairwise TM-scores of generated peptides, with \textbf{Novelty} calculated as the percentage of TM-score less than 0.5. 
\textbf{AAS} refers to the similarity of amino acid sequences between generated and ground truth peptide pairs, quantified as the longest common subsequence ratio. 
Co-analysis with structural novelty can reflect the consistency between sequence and conformation.
% \textbf{AA-Similarity} is quantified by the longest common subsequence ratio between peptide pairs, reflecting the consistency between sequence and conformation.

\subsubsection{Results: Achieves Superior Affinity and Stability}
From the comparison in Table~\ref{co-design}, it can be concluded that 
(i) THFlow exhibits a significant advantage in energy metrics, achieving 6.99\% higher stability and 4.62\% higher affinity compared to baseline models.
A key distinction from flow-based methods explains that this improvement is largely due to the effective handling of multimodal temporal inconsistency problem.
(ii) Our method generates diverse and novel peptide conformations. In contrast, the stronger randomness of diffusion models enables broader exploration of the conformational space, while THFlow focuses on finer adjustments near the pocket, leading to slightly lower diversity.
(iii) The designed sequences also show corresponding differences, reflecting the consistency between sequence and structural in the co-design process.
(iv) Ablation experiments show the positive effects of interaction-related information and sequence-structure balance weight modules on the model. 
The former can deepen the understanding of the physical and chemical significance of the binding interface, and bring the improvement of the affinity. 
The latter shows the importance of the interaction between structure and sequence in training. 
After removing the constraints, the sequence generated by the model is more similar to the ground truth, but the structure is more novel, indicating that there are many unreasonable new structures generated in the same environment, which is reflected in the worse affinity and stability.

\begin{figure*}[!htb]
    \centering
    \includegraphics[width=0.99\linewidth]{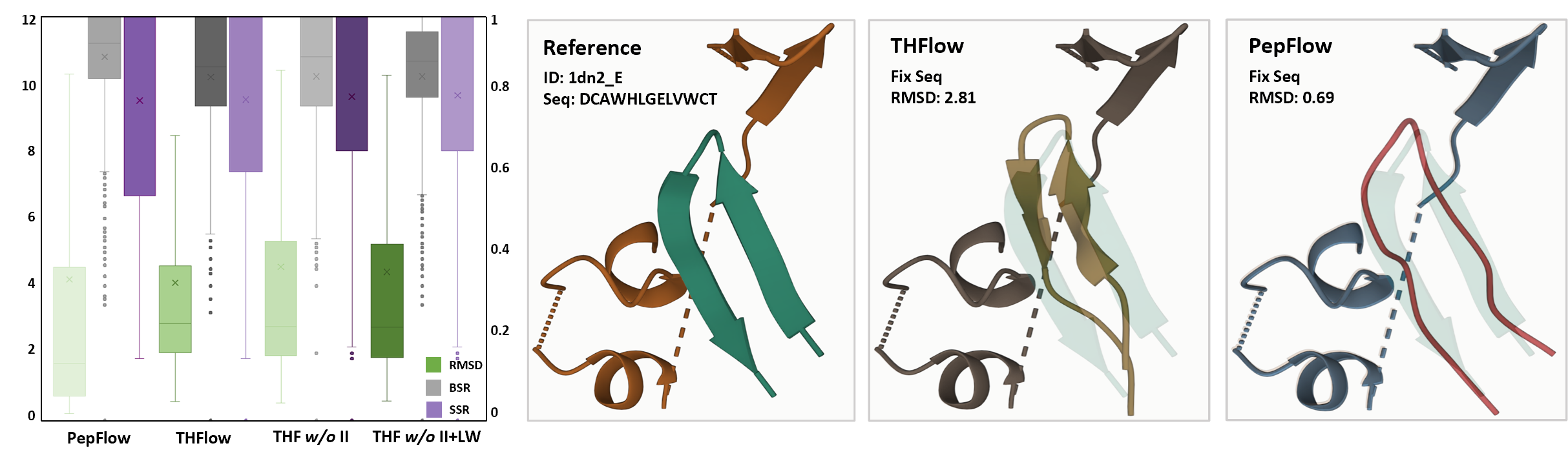}
     % \vspace{-2mm}
    \caption{\textbf{\textit{Left:}} Box plots of metrics for  the re-docking task, including \textbf{lowest RMSD} and \textbf{best binding site similarity}. \textbf{\textit{Right:}} Reference and generated peptide for the re-docking task, highlighting the accurate \textbf{restoration of the peptide structure}.}
    \label{fig:re-dock-result}
    \vspace{-2mm}
\end{figure*}

\subsection{Peptide Re-docking}
The re-docking task evaluates the model's ability to reconstruct the conformation by fixing the reference peptide sequence during the sampling stage, without retraining the model, and generating a full-atom peptide-protein complex in the docked state.
We evaluate two baseline models in re-docking task: \begin{sc}PepFlow\end{sc} and \begin{sc}HDOCK\end{sc}~\cite{yan2020hdock}.
\begin{sc}HDOCK\end{sc} is a traditional docking method, which uses a combination of rigid-body docking followed by energy minimization to predict the binding mode of two interacting proteins. 

\begin{table}[!h]
% \vspace{-2mm}
% \begin{center}
\caption{Comparison for Peptide Re-docking: \textbf{Lowest RMSD}. \textit{Suc.} $=$Success, \textit{Div.} $=$ Diversity}
\label{tab:re-docking}
\centering
% \begin{small}
% \begin{sc}
\begin{tabular}{cccccc}
\toprule
\textbf{Method}      & \textbf{RMSD $\downarrow$} & \textbf{SSR $\uparrow$} & \textbf{BSR $\uparrow$} & \textbf{Suc. $\uparrow$} & \textbf{Div.} \\
\midrule
HDOCK       & 24.16        & 14.02\%  & -  & 12.71\%  &  - \\
PepFlow     & 4.19         & 79.18\%  & \textbf{90.02\%} & 58.89\% & 0.481 \\
\midrule
THFlow      & \textbf{4.09}   & 80.01\%  & 85.48\% & 52.15\% & 0.638 \\
THF \textit{w/o} II  & 4.56  & 80.64\% & 85.67\% & \textbf{60.12\%} & \textbf{0.647} \\
THF \textit{w/o} II+LW  & 4.41  & \textbf{80.91\%} & 85.63\% & 58.89\% & 0.624 \\
\bottomrule
\end{tabular}
% \vspace{-2mm}
\end{table}

\subsubsection{Metrics}
The evaluation of the peptide re-docking task is based on five metrics: 
\textbf{RMSD} measures the structural deviation between the generated peptide's $C_{\alpha}$ atoms and the reference peptide, indicating conformation accuracy;
\textbf{SSR} quantifies the similarity in secondary structure, assessing the preservation of secondary structural features;
\textbf{BSR} calculates the overlap between the docking site of the generated and the reference peptide-protein complex, reflecting binding site accuracy; 
\textbf{Success} is defined as achieving a top-1 \text{RMSD} $<$ 4 \AA, with BSR and SSR both greater than 0.8. The result reports the percentage of successful cases;
and \textbf{Diversity}, same as the co-design task, calculated as 1 minus the average TM-score, measuring the variation in peptide conformations.

\subsubsection{Results: Lower RMSD in Reconstructing}
Table~\ref{tab:re-docking} shows that 
(i) THFlow achieves the lowest RMSD, this demonstrates that the correct docking process simulation provides the model with a strong ability to reconstruct conformations.
(ii) As illustrated in the right of Figure~\ref{fig:re-dock-result}, we found many examples showing THFlow's attempts to understand and reduce the secondary structure of peptides.
Although this may result in a higher RMSD, we consider it as the model exploring the association between coordinates, structure, and interactions. 
This also explains why our model has higher diversity.
% (ii) Compared to flow-based models, the traditional method HDOCK, which does not redesign the peptide structure, does not provide SSR or diversity results, limiting its ability to model peptide-protein interactions. 
% (iii) Among our model variants, removing interaction force information (\textit{w/o} II) increases the success rate, as it shifts the model's focus from balancing structural alignment and docking interactions to solely optimizing structural alignment.
(iii) Removing interaction information (\textit{w/o} II) improves the success rate, likely because it eliminates a difficult trade-off. This allows the model to dedicate its full capacity to optimizing structural alignment instead of balancing it with docking forces.

\subsection{Side-chain Packing}
This task evaluates the model's ability to predict the correct torsional angles for the side chains, which are crucial for accurate protein-ligand docking and stability. 
Specifically, we calculate the mean squared error (MSE) of the four predicted side-chain torsional angles. We use energy-based methods: \begin{sc}RosettaPacker\end{sc}~\cite{leman2020macromolecular}, \begin{sc}SCWRL4\end{sc}~\cite{krivov2009improved}, and Rotamer Density Estimator (RDE)~\cite{luo2023rotamer} with Conditional Flow on TNrt: \begin{sc}RDE-PP\end{sc}~\cite{lin2024ppflow} as baselines.
% \vspace{-2mm}

\begin{table}[!htb]
\vspace{-3mm}
\centering
% \begin{small}
\caption{Comparison for Side-chain Packing}
\label{side-chain}
% \begin{tabular}{lcccc}
\begin{tabular}{cp{0.7cm}p{0.7cm}p{0.7cm}p{0.7cm}}
\toprule
\textbf{Method}      & \textbf{$\chi_1$} & \textbf{$\chi_2$ }& \textbf{$\chi_3$} &\textbf{$\chi_4$} \\
\midrule
Rosetta     & 38.31    & 43.23    & 53.61   & 71.67 \\
SCWRL4      & 30.06    & 40.40    & 49.71   & \textbf{53.79} \\
RDE-PP      & 37.24    & 47.67    & 66.88   & 62.86 \\
PepFlow     & 17.38    & 24.71    & \textbf{33.63}   & 58.49 \\
\midrule
THFlow      & \textbf{11.69} & \textbf{19.66} & 54.87 & 55.74\\
\bottomrule
\end{tabular}
% \end{small}
% \vspace{-2mm}
\end{table}

\subsubsection{Results: Strong Performance in Lower-Order Torsions}
As shown in Table \ref{side-chain}, THFlow demonstrates a clear advantage in predicting the \( \chi_1 \) and \( \chi_2 \) torsional angles, outperforming all other methods. 
This highlights the importance of adjusting torsion angles according to the pocket environment, reflecting THFlow's ability to finely tune the conformation near the pocket.
However, for the more challenging \( \chi_3 \) and \( \chi_4 \) angles, the performance of all models is comparable, with showing worse results compared to \( \chi_1 \) and \( \chi_2 \). 
This highlights that higher-order torsions remain a complex task, with no model consistently outperforming others across all angles.

\subsection{One-step Generation}
Most flow matching methods rely on the assumption that the direction of the gradient vector field remaining constant, which theoretically enables one-step generation. 
However, achieving this in practice is challenging due to two key issues: first, path conflicts as identified by Rectified Flow~\cite{liu2022flow} exist across all flow-based models;
second, our polynomial assumptions hinder the ability of the initial time step to capture temporal inconsistencies across different modalities, limiting the effectiveness of one-step generation.

In this task, we set varying number of inference steps (N) in sampling stage for sequence-structure co-design task to evaluate how it impacts the model’s ability and determine the minimum steps required to capture the temporal inconsistencies in multimodal data.

\begin{figure}[!htbp]
\vspace{-2mm}
\label{fig:step}
    \centering
    \begin{minipage}[b]{0.49\linewidth}
        \centering
        \includegraphics[width=\linewidth]{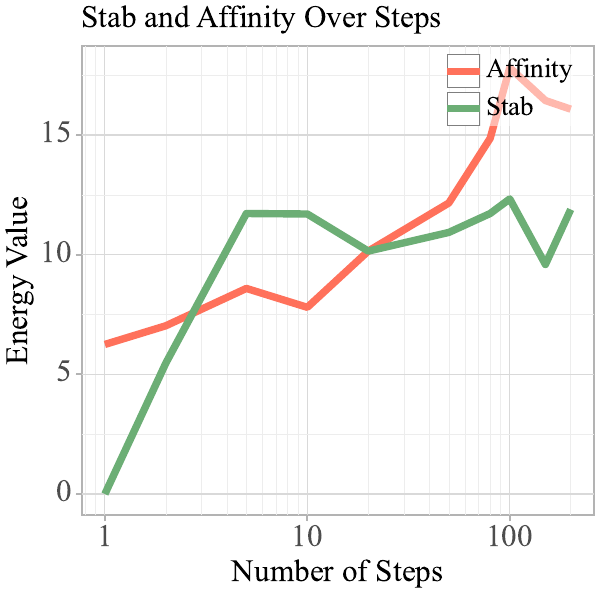}
        % \caption{energy versus number of steps}
    \end{minipage}
    \hfill
    \begin{minipage}[b]{0.49\linewidth}
        \centering
        \includegraphics[width=\linewidth]{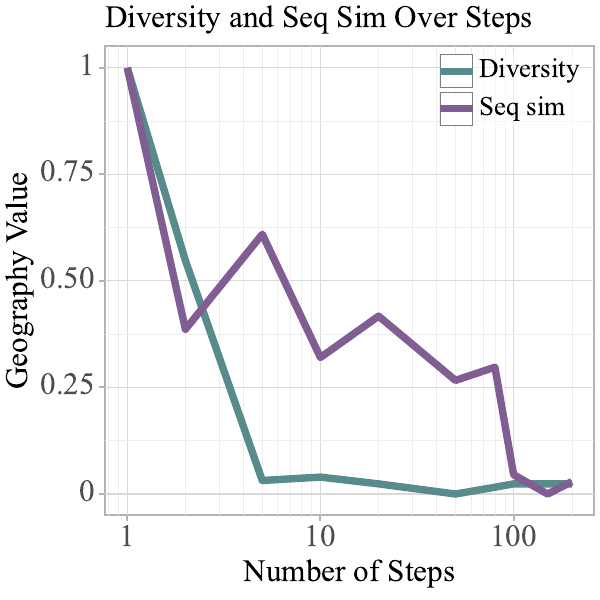}
        % \caption{Image 2}
    \end{minipage}
% \vspace{-4mm}
\caption{Metrics in different inference steps. \textbf{\textit{Left:}} Energy metrics. \textbf{\textit{Right:}} Geography metrics (normalized to range $[0, 1]$).}
\vspace{-2mm}
\end{figure}

\subsubsection{Results: Rapid Improvement with Minimal Steps}
As shown in Fig~\ref{fig:step}, at $step = 1$, the lack of temporal consistency across modalities leads to poor performance, particularly with stability (Stab = 0). However, by $step = 2$, the model rapidly improves, indicating that the difficulty of one-step generation primarily arises from the absence of temporal information, highlighting the importance of the temporally hierarchical assumption. 
After $step > 10$, further increasing the number of steps has minimal impact on the model's performance, suggesting that our model largely follows the straight-line assumption of flow, making it a fast and efficient generation model.

\section{Conclusion}  
In this work, we present \textit{THFlow}, a temporally hierarchical flow matching framework designed to address the critical challenge of temporal inconsistency across modalities in full-atom peptide design. 
By introducing a polynomial-based interpolation scheme and its associated time-varying gradient vector field, THFlow explicitly decouples the evolution of the positional manifold from rotational, torsional, and residue-type manifolds. 
This approach captures the biologically hierarchical nature of peptide docking, resulting in peptides with enhanced binding affinity and stability. 
The integration of interaction-related features further enables the model to balance structural alignment with functional optimization, providing a nuanced simulation of peptide-protein interactions. 

Despite these advancements, THFlow remains limited by suboptimal accuracy in predicting higher-order torsions under complex conformational constraints, 
and its predefined polynomial interpolation, though effective for temporal decoupling, restricts flexibility in modeling dynamic temporal hierarchies. 
Future directions include developing adaptive flow architectures to learn temporal dynamics directly from data, and incorporating dynamic interaction force predictors for context-aware refinement. 

\section*{Acknowledgment}
This work was supported by the National Natural Science Foundation of China (grants No. 62172273), the Science and Technology Commission of Shanghai Municipality (grants No. 24510714300), and the Shanghai Municipal Science and Technology Major Project, China (Grant No. 2021SHZDZX0102).

\bibliographystyle{IEEEtran}
\bibliography{IEEEexample}

\end{document}